\input{aipcheck} 
\documentclass[
final            
] {aipproc}
\usepackage{amsbsy} 
\layoutstyle{6x9} 
\def\ppb{{\bar{\mathrm{p}}\mathrm{p}}}
\def\LLb{{\overline{\Lambda}\Lambda}}
\def\ppLL{{\bar{\mathrm{p}}\mathrm{p}\to\overline{\Lambda}\Lambda}}
\def\vec#1{\boldsymbol{#1}} 
\begin{document} 
\title{Spin observables in
\boldmath$\ppLL$\unboldmath \\  and density-matrix constraints
\footnote{Contribution to LEAP 05, May 16-22, 2005, to appear in the  Proceedings}%
$^{\; ,}$%
\footnote{Preprint \# LPSC 05-70,\qquad ArXiv:nucl-th/0508060}} 
\author{Mokhtar Elchikh}{address={%
D\'epartement de Physique, USTO, BP 1505
El M'naouer, Oran Algeria}} 
 \author{Jean-Marc Richard}{address={%
Universit\'e Joseph Fourier-CNRS-IN2P3, 53, av. des Martyrs, 38026 Grenoble
cedex, France}} 
\begin{abstract} The positivity conditions of the spin
density matrix constrain the spin observables of the reaction $\ppLL$,  leading
to  model-independent,  non-trivial inequalities. The formalism is briefly
presented and examples of inequalities are provided. 
\end{abstract} 
\maketitle 
\section{Introduction} 
The strangeness-exchange reaction $\ppLL$ 
has been studied at low energy by the PS185 collaboration with the antiproton
beam of the LEAR facility at CERN.  Experimental data on spin observables with
a transversely-polarized proton target have been published
\cite{Bassalleck:2002sd,Pas2001}. This contribution is devoted to the 
inequalities relating two or three spin observables, which can be derived
either empirically or by imposing positivity conditions to the  density matrix.
\section{Empirical approach}
 In Ref.\cite{Elchikh:1999ir}, a number of inequalities among the spin
observables has been written down.  The method consists in generating randomly
the real and imaginary parts of the complex amplitudes, computing the various 
observables and plotting one observable against another.  Each observable
$\mathcal{O}_i$ is typically normalized as $-1\le\mathcal{O}_i\le+1$. If a pair
of randomly-generated observables, $\{\mathcal{O}_i,\mathcal{O}_j\}$, covers
the whole square $[-1,+1]^2$, there is no correlation between these
observables. Very often, however, the domain is restricted to a disk or a
triangle inner to the square, revealing that there exists an inequality of the
type $\mathcal{O}_i^2+\mathcal{O}_j^2\le~1$, or  $4\mathcal{O}_j^2\le
(\mathcal{O}_i+1)^2$, which can be derived by inspecting the explicit
expressions of these observables. 

This method can be extended  to the case of a triplet of observables. Examples
of such plots are given in the figure. In the third plot, an
inequality  $\mathcal{O}_1^2+\mathcal{O}_2^2+\mathcal{O}_3^2\le1$ is observed,
with obvious consequences  for the projections,such as
$\mathcal{O}_1^2+\mathcal{O}_2^2\le1$. In the fourth plot, however, the 
$\{\mathcal{O}_1,\mathcal{O}_2,\mathcal{O}_3\}$ domain is limited by a cubic
surface, but there is no restriction on each $\{\mathcal{O}_i,\mathcal{O}_j\}$
pair.

\begin{figure}[!h]
\centerline{%
\includegraphics[width=.25\textwidth]{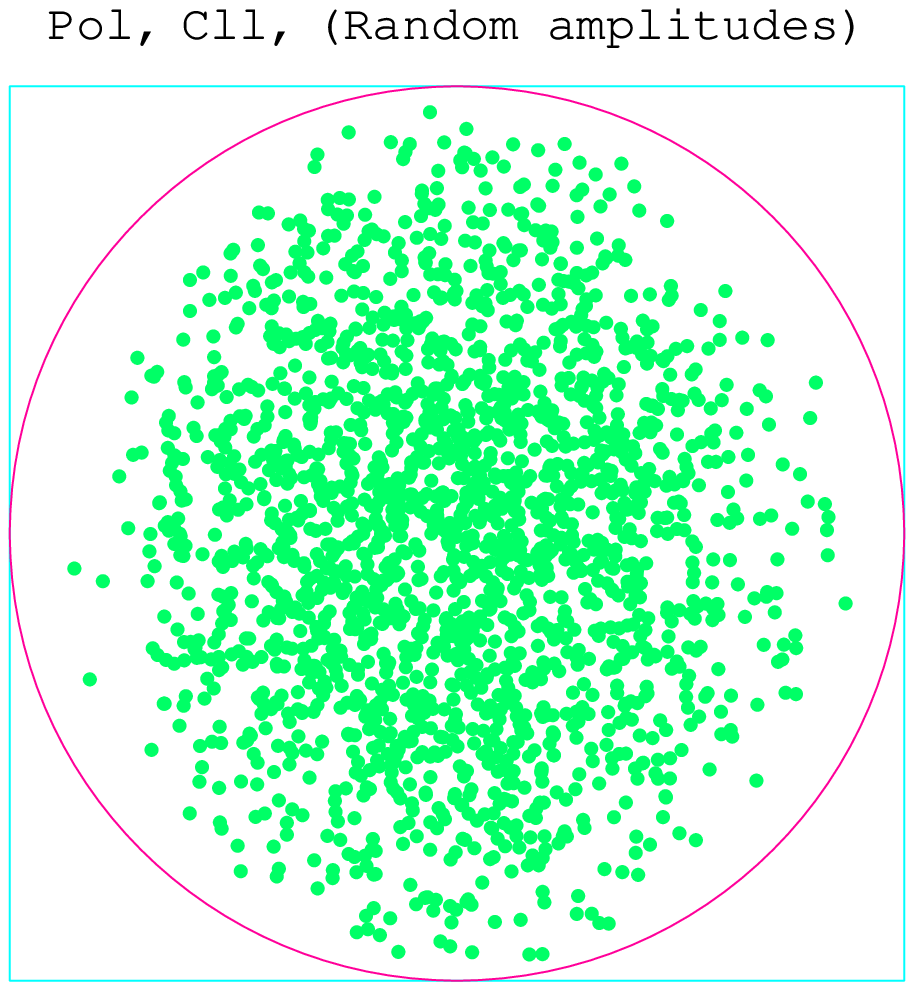}
\includegraphics[width=.25\textwidth]{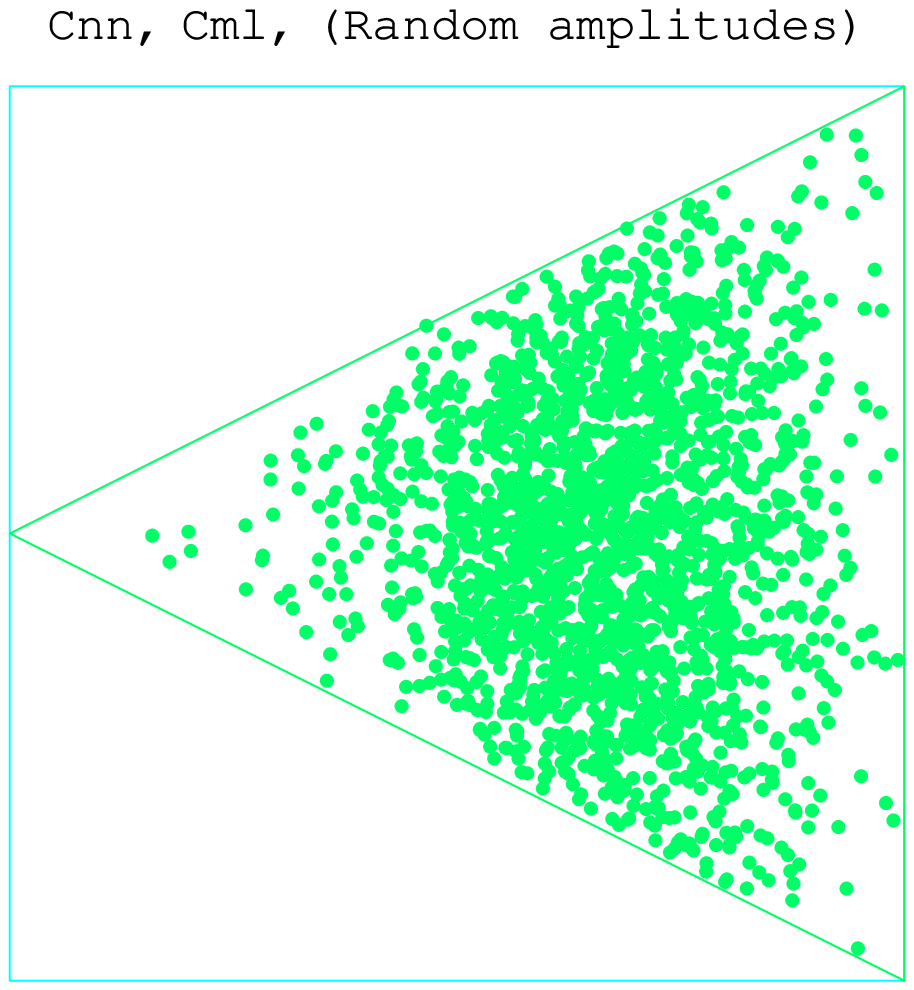}
\includegraphics[width=.25\textwidth]{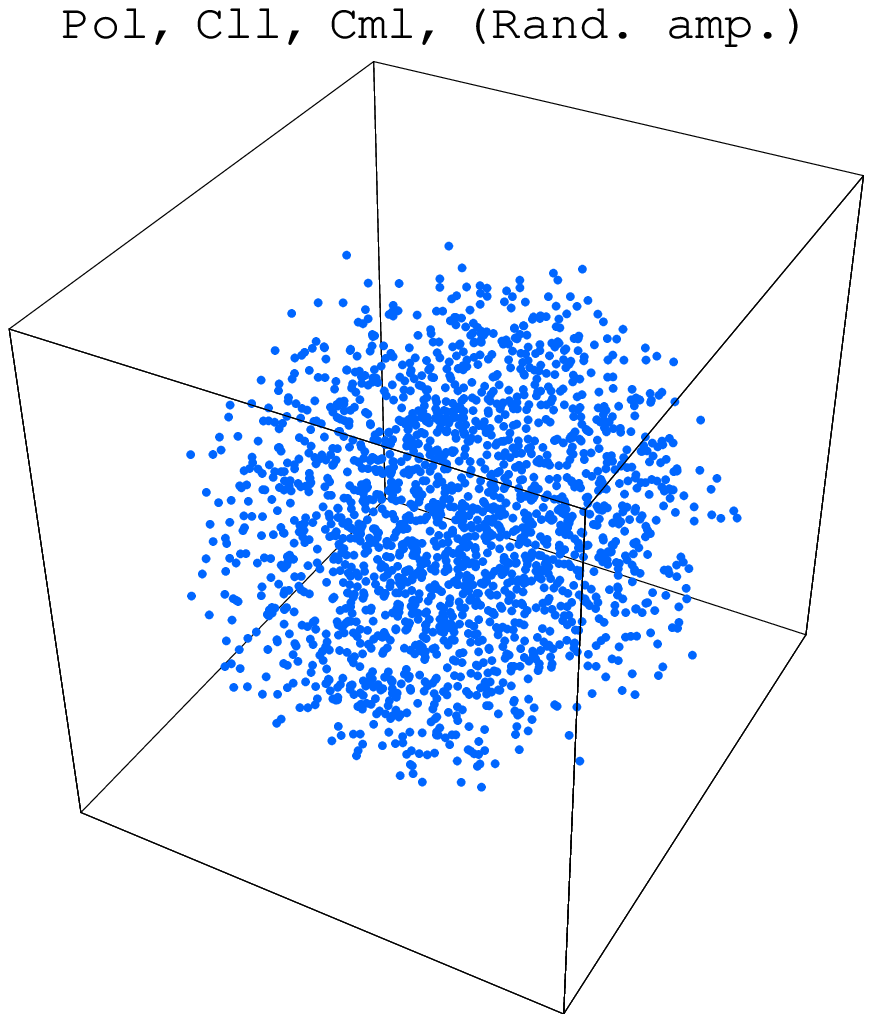}
\includegraphics[width=.25\textwidth]{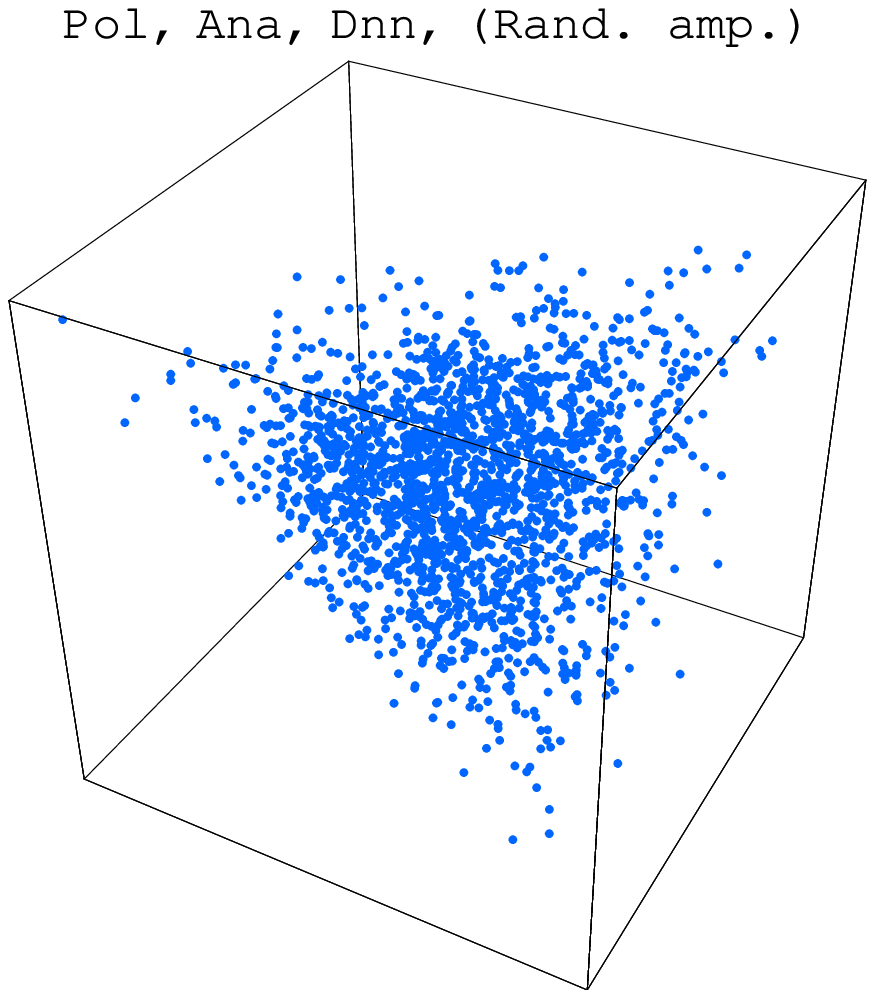} }
\caption{Simulation of  observables by randomly generated amplitudes: from left to right, $P$ vs.\ $C_{ll}$, $C_{nn}$ vs.\ $C_{ml}$, $P$ vs.\ $C_{ll}$ and $C_{ml}$, and $P$ vs.\ $A$ and $D_{nn}$.}
\end{figure}
\section{Explicit density matrix formalism}
Let us now turn to the formalism of the spin-density matrix. Any diagonal element of the density matrix
$\rho$ is positive, \emph{i.e.}, $\rho _{ii} \geq 0$. For any $2\times 2
$ restriction, $\rho_{ii}.\rho _{jj}\geq
\mid \rho _{ij}\mid ^{2}$. This is sufficient for this survey.  More general relations deduced from 
positivity are  discussed in Ref.\cite{Richard:2003bu,Artru:2004jx}.

The density matrix for a polarized set of particles with spin 1/2 is 
\begin{equation}
\label{rhop1}
 \rho_ {p}=\frac{1}{2}(I+\vec{P}.\vec{\sigma })~,
\end{equation}
where $\vec{\sigma }=\{\sigma_1,\sigma_2,\sigma_3\}$ is made of the usual Pauli matrices, and
$\vec{P}$ is the vector polarization. For the proton target, in our case, $\vec{P}$ is transverse  to the
unit-vector $\widehat{z}$ indicating the direction of the antiproton beam. In the ideal case of a 100\% polarised target, 
\begin{equation}
\label{polvector} \vec{P}=\sin{\phi}\ \hat{x}+\cos{\phi}\ \hat{n}~,
\end{equation}
where $\hat{n}$ is normal to the scattering plane, and $\hat{x}=\hat{n}\times\hat{z}$.
The spin-density matrix of the initial $\ppb$ state is thus
\begin{equation}
\label{rhopbarp2} 
\rho_\ppb=\frac{1}{4}[I\otimes I+\sin{\phi}\  (I \otimes
\widehat{x}.\overrightarrow{\sigma })+ \cos{\phi}\ (I \otimes
\widehat{n}.\overrightarrow{\sigma })]~,
\end{equation}
and we shall adopt the usual convention, see Ref.\cite{Elchikh:2004ex},
that whilst the proton spin is projected on the $\{\hat{z},\hat{x},\hat{n}\}$, the antiproton one is writtten in the basis $\{-\hat{z},-\hat{x},-\hat{n}\}$. The explicit form is
\begin{equation} \label{rhopbarpmat}
\rho _{\ppb}=\frac{1}{4}\sum\limits_{i=0,x,n}P_{i}(I\otimes
\sigma_{i})=\frac{1}{4}\left(
\begin{array}{llll}
1 & -ie^{-i\phi } & 0 & 0 \\
ie^{i\phi } & 1 & 0 & 0 \\
0 & 0 & 1 & -ie^{-i\phi } \\
0 & 0 & ie^{i\phi } & 1
\end{array}
\right)~,
\end{equation}
where $P_{0}\equiv 1$, $\sigma_{0}\equiv I$ (the identity matrix).
If $M$ is the transition matrix (amplitude) of the reaction
$\ppLL$, as written, e.g., in \cite{Elchikh:1999ir}, the density matrix of the final state
$\LLb$ reads
\begin{equation}
\label{rholbarl1}
 \rho_{\overline{\Lambda}\Lambda} =M\rho _{\overline{p}p}
M^{\dagger} =\frac{1}{4}\sum\limits_{i=0,x,n}P_{i}\ M \ (I\otimes
\sigma _{i})\ M^{\dagger}~.
\end{equation}
Using the Pauli
matrices of $\overline{\Lambda}$ and $\Lambda$, respectively, it can be decomposed as
 \begin{equation}
\label{rholbarl2}
 \rho_{\overline{\Lambda}\Lambda} =
\frac{1}{4}
I_{0}\sum\limits_{j,k=0,x,n,z}\left[\sum\limits_{i=0,x,n}P_{i}\
O_{ijk} \ (\sigma_{j}\otimes \sigma_{k})\right]~,
\end{equation}
this defining
\begin{itemize}
    \item the differential cross section $I_{0}\equiv (1/4) \rm{Tr} (MM^{\dagger })$,
    \item  the spin observables. $O_{ijk}\equiv \rm{Tr}[M\ (I\otimes \sigma _{i})
\ M^{\dagger}\ (\sigma _{j}\otimes \sigma _{k}] /\rm{Tr}(MM^{\dagger
})$.
\end{itemize}

More explicitly,
 \begin{equation}\label{rhomat}
\rho_{\overline{\Lambda}\Lambda}(\phi)=\frac{1}{4}I_{0}(C_{0}+\cos{\phi}\
C_{n}+\sin{\phi}\ C_{x})~,
\end{equation}
where
 \begin{equation}
C_{0}\equiv \sum\limits_{j,k} O_{0jk}\ (\sigma_{j}\otimes\sigma_{k})~, \quad
C_{x}\equiv \sum\limits_{j,k} O_{xjk}\
(\sigma_{j}\otimes\sigma_{k})~,\quad 
 C_{n}\equiv
\sum\limits_{j,k} O_{njk}\ (\sigma_{j}\otimes\sigma_{k})~,
\end{equation}
The strong interaction responsible for the  $\ppLL$ reaction 
 conserves many discrete symmetries such as
parity and charge conjugation.  Thus some observables vanish or are related to some others.
It remains a set of only $21$ independent observables. For those of rank 1 or 2, $O_{ijk}$ is replaced by the more  familiar notation: $P$ (polarization), $A$
(asymmetry), $C_{jk}$ (correlation), $D_{jk}$ (spin depolarization)
and $K_{jk}$ (spin transfer), leading to %
\[
   C_{0}=\left(
\begin{array}{cccc}
1-C_{zz} & -C_{xz}-iP_{n} & -C_{xz}-iP_{n} & -C_{nn}-C_{xx} \\
-C_{xz}+iP_{n} & 1+C_{zz} & C_{nn}-C_{xx} & C_{xz}-iP_{n} \\
-C_{xz}+iP_{n} & C_{nn}-C_{xx} & 1+C_{zz} & C_{xz}-iP_{n} \\
-C_{nn}-C_{xx} & C_{xz}+iP_{n} & C_{xz}+iP_{n} & 1-C_{zz}
\end{array}
\right)~,
\]
\begin{equation}C_{x}= \left(
\begin{array}{cccc}
-D_{xz}+K_{xz} &-iO_{xzn}-D_{xx} &
K_{xx}+iO_{xnz}& i(O_{xnx}-O_{xxn}) \\
-D_{xx}+iO_{xzn}& D_{xz}+K_{xz} &
i(O_{xnx}+O_{xxn}) & K_{xx}-iO_{xnz}\\
K_{xx}-iO_{xnz}& -i(O_{xnx}+O_{xxn})
 & -D_{xz}-K_{xz} & -D_{xx}+iO_{xzn} \\
i(O_{xxn}-O_{xnx}) & K_{xx}+iO_{xnz} & -D_{xx}-iO_{xzn} &
D_{xz}-K_{xz}
\end{array}\right)~,
\end{equation}
\[C_{n}=\left(
\begin{array}{cccc}
A_{n}+O_{nxx} & -O_{nzx}-iD_{nn} &
-O_{nxz}-iK_{nn} & -A_{n}-O_{nxx} \\
-O_{nzx}+iD_{nn} & A_{n}-O_{nxx}&
A_{n}-O_{nxx} & O_{nxz}-iK_{nn} \\
-O_{nxz}+iK_{nn} & A_{n}-O_{nxx}
& A_{n}-O_{nxx} & O_{nzx}-iD_{nn} \\
-A_{n}-O_{nxx} & O_{nxz}+iK_{nn}
 & O_{nzx}+iD_{nn} & A_{n}+O_{nxx}
\end{array}
\right)~.
\]


The relation $\rho_{11}\rho_{22}\geq|\rho_{12}|^{2}$ gives
\begin{equation}\label{r1}
    (1-C_{zz})(1+C_{zz})\geq \left| -C_{xz}-iP_{n}\right|^{2}~,
\quad\hbox{i.e.,}\quad
C_{xz}^{2}+P_{n}^{2}+C_{zz}^{2}\leq 1~,
\end{equation}
which, of course, implies
\begin{equation}\label{ineq1}
C_{xz}^{2}+P_{n}^{2}\leq 1~,\quad
C_{zz}^{2}+P_{n}^{2}\leq 1~,\quad
C_{xz}^{2}+C_{zz}^{2}\leq 1~.
\end{equation}
Similarly, 
\begin{equation}
(1+C_{zz})^2\geq (P_{n}+C_{xz})^2~.
\end{equation}


If the polarization of the proton target is introduced, 
the positivity of $\rho_{\overline{\Lambda}\Lambda}(0)$ and
$\rho_{\overline{\Lambda}\Lambda}(\pi)$,  which mixes the elements of
the two blocks $C_{0}$ and $C_{n}$, induces
\begin{equation} \label{ineq2}
O_{nxx}^{2}+O_{nzx}^{2}+C_{xz}^{2}+P_{n}^{2}+D_{nn}^{2}+C_{zz}^{2}\leq
1+A_{n}^{2}~.
\end{equation}
Here, with the use of the explicit expressions of the spin
observables in terms of the complex parameters, it can be shown that 
\begin{equation} \label{ineqCn} D_{nn}^{2}+C_{zz}^{2}+O_{nzx}^2\leq 1~,
 \end{equation}
 which leads to:
$D_{nn}^{2}+O_{nzx}^2\leq 1$ and $C_{zz}^{2}+O_{nzx}^2\leq 1$ and
the already-published \cite{Elchikh:1999ir,Richard:1996bb} inequality: 
\begin{equation}\label{ineq4}
D_{nn}^{2}+C_{zz}^{2}\leq 1~.
\end{equation}

\section{Summary} 
Inequalities among spin observables can be derived either
from the explicit expressions of these observables in terms of the amplitudes,
or from the general properties of the  spin density matrix. These inequalities
provide model-independent test of the data on spin observables. Similar
inequalities can be written down  in the case of inclusive reactions or
spin-dependent parton densities~\cite{Artru:2004jx,Soffer:2003qj}. This will be
the subject of a forthcoming review article \cite{AERST}. The formalism of the
spin density matrix is clearly more powerful, and it suggests a more physical
interpretation of the inequalities, which can be read as the flow of quantum
information from the initial to the final state.


\end{document}